\documentclass[final,authoryear,5p]{elsarticle}

\usepackage{epsfig}

\usepackage{amssymb}

\usepackage[ps2pdf,%
a4paper=true,%
breaklinks=true,%
colorlinks=true,%
pdfauthor={G.\ Rauw \& Y.\ Naz\'e},%
pdftitle={X-ray emission from interacting massive binaries}%
]{hyperref}

\journal{Advances in Space Research}

\begin{document}
\newcommand{\lxlb}{$L_{\rm X}/L_{\rm bol}$}
\newcommand{\loglxlb}{$\log[L_{\rm X}/L_{\rm bol}]$}
\newcommand{\aap}{A\&A}
\newcommand{\apj}{ApJ}
\newcommand{\apjl}{ApJL}
\newcommand{\apjs}{ApJS}
\newcommand{\mnras}{MNRAS}
\newcommand{\apss}{Ap\&SS}
\newcommand{\sovast}{Soviet Astronomy}
\newcommand{\aapr}{A\&ARv}
\newcommand{\na}{New Astronomy}
\newcommand{\ssr}{Space Science Reviews}

%%%%%%%%%%%%%%%%%%%%%%%%%%%%%%%%%%%%%%%%%%%%%%%%%%%%%%%%%%%%%%%%%%%%%%%%%%%%%
%% Frontmatter
\begin{frontmatter}


% Use the tnoteref command within \title and fnref within \author or \address for footnotes;
% use the corref command within \author for corresponding author footnotes;
% \author{Name\corref{cor1}\fnref{label2}}
% \address{Address\fnref{label3}}

\title{X-ray emission from interacting wind massive binaries: a review of 15 years of progress}

% Use optional labels to link authors explicitly to addresses:

\author{Gregor Rauw\corref{cor}}
\ead{rauw@astro.ulg.ac.be}
\cortext[cor]{Corresponding author}
\author{Ya\"el Naz\'e\fnref{footnote4}}
\address{Groupe d'Astrophysique des Hautes Energies, Institut d'Astrophysique et de G\'eophysique, Universit\'e de Li\`ege, All\'ee du 6 Ao\^ut, 19c, B\^at B5c, 4000 Li\`ege, Belgium}
\fntext[footnote4]{Research Associate FRS-FNRS (Belgium)}
\ead{naze@astro.ulg.ac.be}

\begin{abstract}
Previous generations of X-ray observatories revealed a group of massive binaries that were relatively bright X-ray emitters. This was attributed to emission of shock-heated plasma in the wind-wind interaction zone located between the stars. With the advent of the current generation of X-ray observatories, the phenomenon could be studied in much more detail. In this review, we highlight the progress that has been achieved in our understanding of the phenomenon over the last 15 years, both on theoretical and observational grounds. All these studies have paved the way for future investigations using the next generation of X-ray satellites that will provide crucial information on the X-ray emission formed in the innermost part of the wind-wind interaction.
\end{abstract}


\end{frontmatter}

\parindent=0.5 cm

%%%%%%%%%%%%%%%%%%%%%%%%%%%%%%%%%%%%%%%%%%%%%%%%%%%%%%%%%%%%%%%%%%%%%%%%%%%%%
%% Main text
\section{Introduction}
Early-type stars of spectral type OB or Wolf-Rayet (WR) feature powerful, highly supersonic stellar winds that are driven by the radiation pressure in numerous spectral lines. These winds combine large mass-loss rates and wind velocities and thus they vehicle important quantities of kinetic power. When two such stars are bound together by gravity in a binary system, their winds interact and part of the kinetic energy is converted into heat. This interaction gives rise to a number of observational signatures that span a wide range of the electromagnetic spectrum, from radio waves to the $\gamma$-ray domain. In this review, we focus on one of the most spectacular observational consequences of colliding winds: the X-ray emission that is produced by the shock-heated plasma in the wind interaction zone. 
 
Two years before the first detection of X-ray emission from massive stars, \citet{PU} and \citet{Chere2} proposed that collisions between the wind of one component of a massive binary system and either of the wind, photosphere or magnetosphere of the companion star should produce strong X-ray emission. Observations with {\it EINSTEIN}, and lateron with {\it ROSAT}, revealed that single and binary early-type stars are moderate X-ray emitters. However, for the binary systems, the observed flux level turned out to be significantly lower than expected from early theoretical considerations. Still, it was found that many of the brighter X-ray sources among massive stars were binary systems, and that, on average, massive binary systems had a larger $L_{\rm X}/L_{\rm bol}$ ratio than single massive stars. This conclusion held for binaries hosting Wolf-Rayet stars \citep{Pollock1} as well as for systems of spectral-type O + O \citep{CG}. Moreover, the level of X-ray emission from some of these systems was found to be variable \citep{Corcoran}. These observational results were interpreted in terms of the collision of the stellar winds and led to the general paradigm that colliding wind binaries should be X-ray brighter than single massive stars. 

This rather simple picture has been refined quite substantially over the last two decades as more detailed observations and more sophisticated models became available. In the following we will summarize the progress that has been achieved in understanding the phenomenon both theoretically (Sect.\,\ref{theory}) and from an observational point of view (Sect.\,\ref{observation}). We deliberately leave $\eta$~Car aside as this system is subject of a dedicated review in this volume (Corcoran et al.). In Sect.\,\ref{conclusion}, we highlight our conclusion and discuss some future perspectives. 

\section{Theoretical models of wind-wind interactions \label{theory}}
Following the observational detection of X-ray emission from colliding winds at a flux level significantly below the anticipated one, major efforts have been devoted to improve theoretical models of wind-wind interactions. Whilst some analytical or semi-analytical formalisms have been proposed, most of the work has focused on the development of hydrodynamic simulations. Over recent years, the progresses in numerical techniques and computational resources allowed for numerical simulations of growing sophistication that account for an increasing number of physical processes. In this section, we first present some general considerations, before we discuss the more recent developments that resulted in the current theoretical picture of wind-wind interactions. 
\subsection{The basic picture}
The wind interaction zone is contained between two oppositely faced shocks separated by a contact discontinuity. The location and shape of this contact discontinuity are mainly determined by the value of the wind momentum ratio \citep[e.g.][]{Shore,Luo,SBP,Usov,Canto}:
\begin{equation}
    \label{eq:1}
\eta = \frac{\dot{M}_1\,v_1}{\dot{M}_2\,v_2}
  \end{equation}
where $v_j$ and $\dot{M}_j$ stand for the pre-shock velocity and mass-loss rate of star $j$. To first approximation, the inner part of the wind interaction zone can be described as a shock cone wrapped around the star with the weaker wind. A popular method to compute the shape of the contact discontinuity is the analytical formalism proposed by \citet{Canto}. These authors study the properties of the wind interaction zone for winds colliding at their terminal velocities assuming that the wind interaction zone forms a thin shell around the contact discontinuity. Using first principles, \citet{Canto} provide non-linear equations for the shape of the contact discontinuity, as well as expressions for the mass surface density of the wind interaction zone and the tangential velocity along the contact discontinuity.\\

At the shock front, the kinetic energy of the inflowing wind normal to the shock is transformed into thermal energy. Given the typical velocities of the stellar winds of early-type stars, the shocked material is heated to temperatures of several $10^7$\,K. What happens to the shock-heated plasma depends strongly on the efficiency of radiative cooling. \citet{SBP} propose to quantify the efficiency of this process via the parameter   
\begin{equation}
    \label{eq:2}
\chi = \frac{t_{\rm cool}}{t_{\rm esc}} = \frac{v^4\,d}{\dot{M}}
  \end{equation}
where $t_{\rm cool}$ and $t_{\rm esc}$ stand for the cooling and escape time, respectively\footnote{Equation\,\ref{eq:2} assumes that the cooling function is independent of the temperature which is only the case over a limited temperature range near $10^7$\,K, see e.g.\ Fig.\,10 of \citet{SBP}.}. Here $v$, $d$ and $\dot{M}$ indicate the pre-shock velocity (in 1000\,km\,s$^{-1}$), the separation between the star and the shock (in $10^7$\,km) and the wind mass-loss rate (in $10^{-7}$\,M$_{\odot}$\,yr$^{-1}$), respectively. Each of the two winds has its own value of $\chi$ \citep{SBP}. For situations where $\chi << 1$, the cooling is very fast and the shock-heated material instantaneously radiates its energy away so that the wind is essentially isothermal. Conversely, if $\chi \geq 1$, the shocked plasma is adiabatic. In this case, radiative cooling is absent and the shock-heated plasma in the wind interaction zone only cools through adiabatic expansion. 

\citet{Luo} presented the first hydrodynamic simulation of colliding winds. These authors focused on two extreme situations: a purely radiative and a purely adiabatic system. The hydrodynamic simulations of \citet{SBP} provided a major step forward as they were the first to explicitly include radiative cooling in a self-consistent way, therefore allowing to consistently treat systems that are partially radiative.  

An important limitation of these early simulations concerns their spatial resolution in the treatment of radiative wind interaction zones. Indeed, whilst adiabatic shocks are self-similar and simply scale with the separation between the stars, the situation is more complex when radiative cooling is important. In the latter case, the cooling length introduces another scale length in the problem \citep{SBP}. Dealing with a spatial grid that fully resolves the structures of radiative shocks whilst simultaneously encompassing the whole wind interaction zone is very challenging in terms of numerical resources. Early hydrodynamic simulations thus failed to resolve the cooling layers. Nowadays the situation has improved thanks to the use of adaptive mesh refinement algorithms \citep[e.g.][and references therein]{Parkin14}, although highly radiative shocks remain a major challenge for numerical simulations. These difficulties of the hydrodynamic simulations to cope with radiative shocks motivated the development of a semianalytic formalism for computing the X-ray emission from such shocks \citep{Antokhin04}. In this approach, the wind acceleration is treated via the wind acceleration laws of single stars. The wind interaction is idealized as an axisymmetric, laminar, steady radiative shock and the cooling layer behind the shock is assumed to be isobaric. Since mixing of hot and cool material in the post-shock region is ignored in this model, it provides an upper limit on the actual level of X-ray emission.\\

An important question is indeed how much X-ray emission is produced by the shock-heated plasma. \citet{Usov} presented a number of analytical recipes to predict the X-ray luminosities of winds colliding at $v_{\infty}$. When compared to actual observations, this formalism was frequently found to overpredict the level of the X-ray flux. \citet{SBP} showed that in the adiabatic case, one expects the X-ray luminosity to approximately scale as 
\begin{equation}
    \label{eq:3}
L_X \propto \frac{\dot{M}^2\,v^{-3.2}}{d}\,(\eta^{-2} + \eta^{-3/2}) 
  \end{equation}
where the symbols have the same meaning as before.
Conversely, for the purely radiative case, \citet{SBP} found that 
\begin{equation}
    \label{eq:4}
L_X \propto f\,\dot{M}\,v^2 
  \end{equation}
where $f$ is the fraction of the wind involved in the collision.
 
The question of which shocked wind is responsible for most of the X-ray emission in colliding wind interactions was re-addressed by \citet{PS}. They showed that in situations where both winds cool rapidly, the faster wind is generally the one that dominates the X-ray emission. Conversely, in adiabatic systems, the stronger wind (i.e.\ the one with the larger wind momentum, $\dot{M}\,v$) is usually the one that emits most of the X-rays. The conclusions of \citet{PS} largely disagree with the recipes of \citet{Usov}. In general, both the analytic and hydrodynamic calculations from these early models predict X-ray luminosities that are an order of magnitude too large compared to the observed values. \citet{SBP} also pointed out that their models for V\,444~Cyg (WN5 + O6, $e = 0$, $P_{\rm orb} = 4.2$\,d) predict a plasma temperature well above that measured on {\it EINSTEIN} spectra.

\subsection{The physics of the shocks \label{shockphysics}}
Our current picture of hydrodynamic shocks in colliding wind binaries is significantly more complex than a simple application of the Rankine-Hugoniot conditions in the limit of strong shocks. Additional effects need to be taken into account. These concern modifications of the properties of the pre-shock flow as well as microscopic effects in the post-shock plasma.\\ 

In a massive binary system, the stellar wind material interacts with the radiation fields and the gravity of both stars. As a result, applying the stellar wind formalism of single massive stars to the winds in binary systems only provides a first order approximation. \citet{SP} were the first to investigate this problem along the axis between the stars in close massive binary systems. They found that the radiation field of the companion inhibits the acceleration of one star's wind preventing it from reaching $v_{\infty}$. Whilst the gravitational field of the companion compensates part of the effect, it cannot prevent a lowering of the pre-shock velocity. Radiative inhibition therefore lowers the post-shock temperature and \citet{SP} suggest that this effect could explain why the X-ray emission from close massive binaries is significantly softer than expected from the terminal velocities of the stellar winds.

\citet{Gayley} considered another consequence of the radiation pressure of the companion. This effect, called radiative braking, is most relevant for systems that exhibit a large imbalance between the wind momenta, such as WR + O binaries. In such systems, the dominant wind has reached an asymptotic velocity, but is suddenly decelerated by the radiation it encounters upon approaching the companion. Radiative braking could lead to a wider opening angle of the shock cone (e.g.\ in V\,444~Cyg, Sect.\,\ref{opacity}). The same effect could prevent a collision of the dominant wind onto the companion's surface in situations where the companion's wind is unable to sustain the momentum of the incoming wind (e.g.\ at periastron passage in $\gamma^2$\,Vel, WC8 + O7.5\,III/9I, $e = 0.33$, $P_{\rm orb} = 79$\,d). \citet{Gayley} provide a semi-analytical recipe to estimate whether or not radiative braking is important for a given system.

Another way the radiative acceleration of the pre-shock wind could be altered was studied by \citet{ParkinSim}. These authors introduce the concept of self-regulated shocks: the X-ray radiation from the shocked material ionizes the wind material ahead of the shock, thereby suppressing the line force and inhibiting the wind acceleration. \citet{ParkinSim} find that this effect can lead to a lower pre-shock velocity, a correspondingly lower temperature of the post-shock plasma, and enhanced cooling, possibly also rendering the shock front unstable. Yet the intrinsic X-ray luminosity of the shocks was found to remain largely unaltered. For systems with very different wind momenta, \citet{ParkinSim} find that this effect renders radiative braking less effective and significantly extends the range of binary separations over which a wind-photosphere collision is likely to occur.

\citet{MZ} considered the effect of thermal conduction due to electrons on the properties of the shocks in an adiabatic wind interaction zone. According to their results, thermal conduction leads to preheating zones in front of the shocks where the temperature is higher than expected. This in turn leads to lower Mach numbers of the inflowing gas. \citet{MZ} concluded that the higher the efficiency of thermal conduction, the lower the temperature of the shocked gas because the energy is distributed over a larger quantity of material. They predict that the resulting X-ray spectrum should be softer than in the absence of conduction. This thermal conduction is not to be confused with numerical heat conduction in hydrodynamical simulations. \citet{ParkPitt2} showed that the dynamics of the hot gas and the corresponding X-ray emission are affected by this purely numerical problem to an extent that depends on the density and temperature contrast across the contact discontinuity. The strongest effects were found when the shocked wind of one star behaves quasi-adiabatically whilst the shocked wind of the other star is strongly radiative. Higher resolution simulations reduce the artificial heat transfer, but increasing the resolution is not always practicable. 

Energy losses via Comptonization, i.e.\ the redistribution of energy from the thermal electrons in the shocked plasma to the photons of the stellar radiation field following inverse Compton scattering, also affect the properties of the wind interaction region. \citet{MZ1,MZ2} showed that energy losses through Comptonization lead to an increase of the density and a decrease of the temperature of the gas in the interaction zone. Therefore, whilst the emission measure of the shocked gas increases, the peak energy of its emission shifts towards lower energies. As a result, X-ray emission gets softer and thus more sensitive to absorption by the cool wind material.  
  
Finally, in colliding wind binary systems where relativistic particles are accelerated through diffusive shock acceleration (see Sect.\,\ref{nonthermal}), the pre-shock velocity can also be modified through so-called shock modification \citep{PD}. This effect results from the diffusion of non-thermal ions upstream of the shock. These relativistic ions then exert a back-pressure on the pre-shock flow leading to a decrease of the pre-shock velocity and the formation of a so-called shock precursor \citep{PD}. The reduced velocity jump leads to a softer X-ray spectrum. In eccentric binary systems, the degree of shock modification could be dependent on the orbital phase.\\

In the strong shock limit, the kinetic energy normal to the shock is converted into heat according to \citep[e.g.\ ][]{SBP}
\begin{equation}
    \label{eq:5}
k\,T_s = \frac{3}{16}\,m_p\,v_{\perp}^2 
  \end{equation}
where $T_s$, $m_p$ and $v_{\perp}$ are the post-shock temperature, the particle mass and the particle's pre-shock velocity perpendicular to the shock, respectively. Because of the very different particle masses, electrons and ions are thus expected to display quite different post-shock temperatures. As pointed out by several authors \citep[][amongst others]{Usov,ZS,Zhekov,Pollock}, in wide binary systems, the postshock densities are such that the timescale for ion and electron temperature equalization through Coulomb interactions are comparable to the escape times. Therefore, one would expect an electron temperature much lower than the ion temperature. Yet, the observed values of the electron temperature are not that low, but rather fall in the range between 0.1 and 1.0 times the ion temperature\footnote{As an example, \citet{Pittard09} assumed in his models that the electron temperature immediately behind the shock amounts to 0.2 times the mean plasma temperature.}. This led to the conclusion that the shocks must be collisionless \citep{ZS,Zhekov,Pollock}. However, the understanding of this phenomenon in the context of colliding wind binaries is still fragmentary, and no clear rule how to deal with this effect has emerged so far. For instance, \cite{PD} suggest that shocks that are efficient in accelerating particles to relativistic energies might equilibrate the electron and ion temperatures much faster than shocks where no such acceleration takes place.  

In a similar context, many observational studies and early models assumed that the shock-heated plasma is in ionization equilibrium. However, in a some cases the timescales needed to reach ionization equilibrium through collisions are longer than the dynamical timescales and the plasma is thus likely not in equilibrium. \citet{Zhekov} proposed to quantify the importance of non-equilibrium ionization effects via the criterion
\begin{equation}
    \label{eq:6}
\Gamma_{\rm NEI} = 1.21\,\frac{\chi_e\,\dot{M}}{\overline{\mu}\,v^2\,d} 
  \end{equation} 
Here, $\chi_e$ is the ratio between the electron and ion number densities, $\overline{\mu}$ is the mean atomic weight for ions, $d$ is the distance from the star to the shock in $10^{11}$\,km, $\dot{M}$ is the mass-loss rate in $10^{-5}$\,M$_{\odot}$\,yr$^{-1}$ and $v$ the wind velocity in units 1000\,km\,s$^{-1}$. For $\Gamma_{\rm NEI} \gg 1$, non-equilibrium effects should play a minor role, whereas they are expected to be important if $\Gamma_{\rm NEI} \leq 1$ \citep{Zhekov}. 

\subsection{The impact of orbital motion \label{Coriolis}}
Since the first 2-D models of colliding winds which neglected the orbital motion and assumed an axisymmetric wind interaction region, there has been enormous progress in the treatment of the impact of orbital motion on the wind-wind interaction. Currently, state-of-the art models use 3-D adaptive mesh refinement hydrodynamical codes accounting for radiative driving, gravity, radiative cooling and orbital motion \citep[e.g.\ ][]{ParkinetaCar,Parkin14}. \\
  
Pioneering work on this subject was performed by the Z\"urich group \citep[e.g.][]{Walder1,Walder,Folini} and by \citet{PS3}. For instance, \citet{Walder} presented a 3-D hydrodynamic simulation of $\gamma^2$~Vel performed with the goal to reproduce the observed {\it ROSAT} lightcurve of the system \citep{Willis}. Although the models of \citet{Walder} assumed a purely adiabatic interaction, whereas the real one is at least partially radiative, the agreement between the simulated and observed lightcurves was rather satisfactory. In these models the unperturbed winds were assumed to have a constant velocity equal to $v_{\infty}$. This assumption was still used in more recent work \citep[e.g.][]{Lemaster}.

In the same context, \citet{Okazaki} performed 3-D smoothed particle hydrodynamics simulations to model the {\it RXTE} X-ray lightcurve of $\eta$\,Car ($P_{\rm orb} = 5.54$\,yr, $e \sim 0.9$). \citet{Okazaki} assumed that the bulk of the X-ray emission arises at the apex of the shock cone. Their models then provided the density structure of the wind interaction zone, hence allowing to compute the associated variations of the column density along the line-of-sight towards the apex. 
 
At the same time, \citet{ParkPitt} presented a simple 3-D dynamical model to evaluate the orbital-induced curvature of the colliding wind region. In this work, the winds were again assumed to collide at terminal velocity. The apex of the colliding wind region was computed adopting the formalism of \citet{Canto}, whilst the downstream contact discontinuity was obtained assuming a ballistic flow. The aberration of the apex due to the orbital motion was approximated by $\tan{\mu} = \frac{v_{\rm orb}}{v_{\infty}}$ (with $v_{\infty}$ the slower of the two wind velocities). \citet{ParkPitt} used this model mainly to estimate the X-ray attenuation by the circumstellar material. The same model was subsequently applied by \citet{Parkin09} to model the {\it RXTE} lightcurve of $\eta$ Car. In contrast with the work of \citet{Okazaki}, \citet{Parkin09} accounted for the spatial extension of the X-ray emission zone. Still, the model failed to reproduce the broad minimum observed in the lightcurve. \citet{Parkin09} accordingly suggested that the secondary wind suffers radiative inhibition at periastron, implying a lower wind velocity and hence stronger post-shock cooling. As a result, the wind collision zone could collapse onto the secondary (see also the review by Corcoran et al., this volume).

\citet{Pittard09} presented 3-D hydrodynamical models, incorporating gravity, the driving of the winds (including radiative inhibition), the orbital motion of the stars and radiative cooling of the shocked plasma. He applied these models to a series of O + O case studies. These simulations revealed some interesting features. For instance, \citet{Pittard09} showed that, in systems with equally strong winds ($\eta = 1$), the gas near the leading shocks is less dense and cooler than the gas near the trailing shocks. This is because, at a given downstream distance from the binary axis, the leading shock is more oblique than the trailing shock. 

Finally, \citet{Lamberts2} considered the structure of an adiabatic colliding wind region on scales much larger than the orbital separation. From first principles, the orbital motion is expected to turn the shock structure into a spiral. \citet{Lamberts2} used an adaptive mesh refinement technique to simulate the large scales whilst simultaneously keeping a high resolution close to the binary. They found that Kelvin-Helmholtz instabilities are triggered by the effect of the orbital motion on the adiabatic wind interaction zone even for winds of equal strength. In some cases, i.e.\ when there exists a steep velocity gradient between the winds, the growth of this instabilities prevents the formation of a stable spiral structure \citep{Lamberts2}. 

\begin{figure*}
\begin{center}
\includegraphics*[width=0.8\textwidth,angle=0]{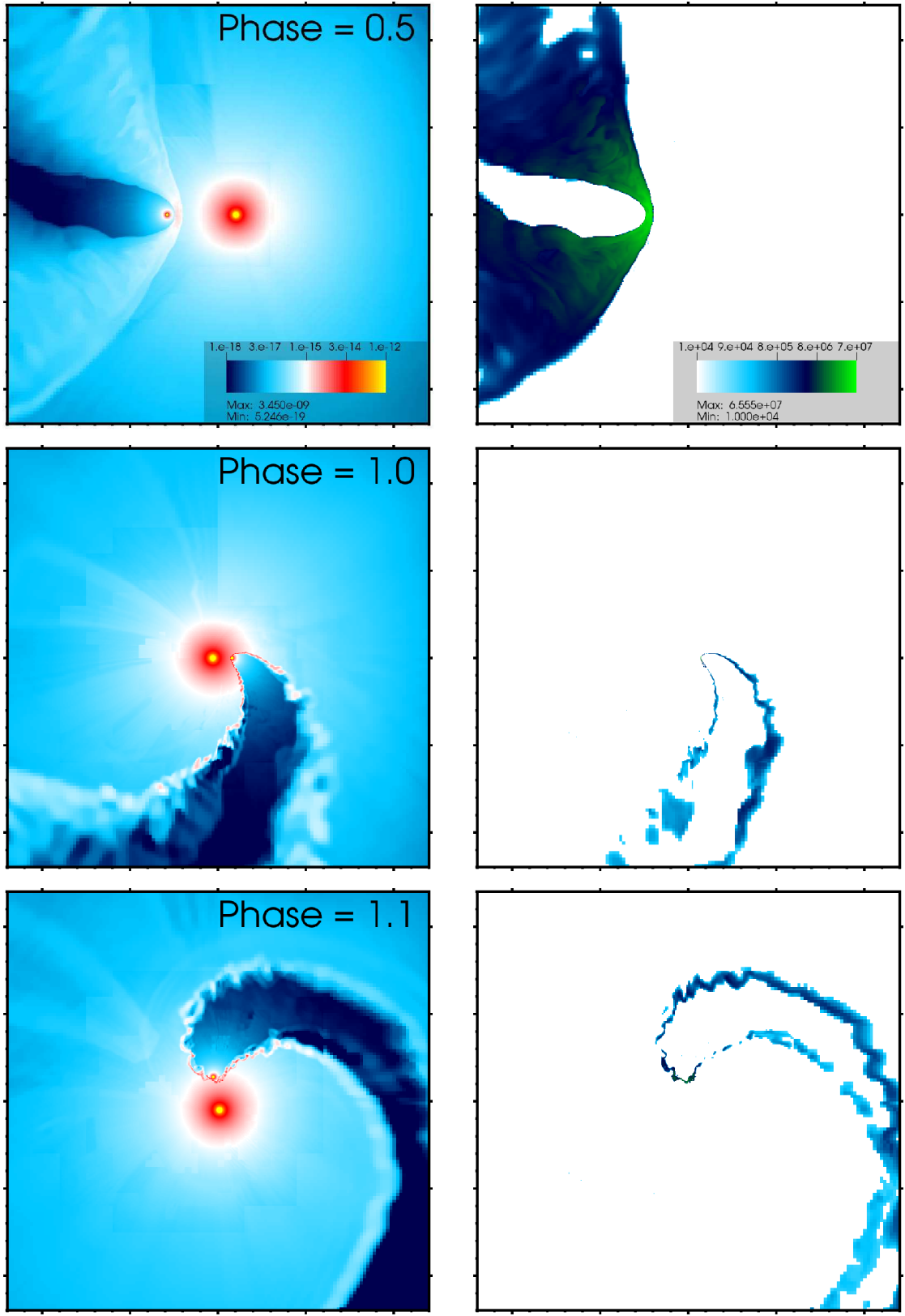}
\end{center}
\caption{3-D hydrodynamic simulation of the wind-wind interaction in WR~22 including the effect of radiative acceleration \citep[from][]{PG}. Left: gas density in the orbital plane at three different orbital phases. Right: temperature of the gas in the orbital plane. The WR star is the star with the stronger wind. Each plot covers a region of $1.2 \times10^{14}$\,cm on one side. Credit: Parkin, E.R. , \& Gosset, E., A\&A, 530, A119, 2011, reproduced with permission \copyright ESO. \label{figure1}}
\end{figure*}

\subsection{The effects of instabilities and clumps \label{theor:instabilities}}
\citet{SBP} drew attention to the importance of dynamical instabilities in their hydrodynamic simulations. They reported the presence of Kelvin-Helmholtz instabilities which arise from the shear when the velocities of the colliding winds are different \citep[see also][]{Lamberts1}. \citet{SBP} also reported on thin-shell instabilities which form when radiative cooling produces shells of cold dense gas behind the shock. 

An important question is of course whether or not these instabilities and their growth reflect a physical phenomenon or stem from limitations of the numerical scheme or specific model assumptions. This aspect was addressed by \citet{MZB} for systems with highly radiative wind interaction zones. In simulations of such systems, its is frequently assumed that the energy provided by stellar radiation prevents the shocked material from having its temperature drop below some threshold value. However, this threshold value is usually not derived self-consistently. \citet{MZB} therefore performed numerical simulations with three different values of this threshold temperature: $10^6$, $10^5$ and $10^4$\,K. These authors found that the choice of the adopted value had a strong impact on the development of the instabilities in the wind-wind interaction zone.

In radiative wind-wind interactions, the dynamical instabilities determine the shape and size of the wind interaction zone as well as its small-scale structure \citep{WalderFolini}. \citet{SBP} predicted that these turbulent motions could lead to variability of the X-ray flux of order 10\%. But these instabilities could also have other far-reaching consequences. Indeed, \citet{Kee} noted that the thin-shell instabilities transform the compressive shock into highly oblique shocks, thus reducing the heating of the material and therefore the associated X-ray emission. These authors presented a parameter study of such instabilities showing that in the radiative regime, the X-ray emission follows indeed a linear scaling with mass-loss, as expected from Equation\,(\ref{eq:4}),  but reduced by about a factor 1/50 compared to analytic radiative-shock models without instabilities \citep[similar to the formalism of][]{Antokhin04}. Though the situations studied by \citet{Kee} consistently yield this factor 1/50 reduction, the universality of this reduction factor and/or its dependence on key parameters such as the Mach number of the pre-shock flow still need to be established.\\

There is nowadays strong observational evidence that the stellar winds of massive stars are highly clumped \citep[e.g.][]{Eversberg}. \citet{che90} tried to explain the low luminosity of colliding wind systems compared to theoretical models, by the fact that only the smooth inter-clump wind is affected by the collision zone between the stars, the clumps passing unaffected through it (and later colliding onto the companion, without X-ray production). \citet{AB} argued that the shocked clumps should be highly radiative, thereby counterbalancing the $1/d$ trend expected for the adiabatic behaviour of the lower density smooth wind component in eccentric systems such as WR~140 (WC7 + O5.5, $e = 0.89$, $P_{\rm orb} = 7.93$\,yr). However, these conclusions are not supported by the hydrodynamic simulations of the collision between clumpy winds in WR~140 presented by \citet{Pittard07}. In the latter work, it was shown that, in situations where the wind interaction zone is adiabatic\footnote{So far, no study of the influence of clumps on radiative wind interaction zones has been presented.}, the clumps are rapidly destroyed in the post-shock region. This clump destruction occurs mainly through dynamical instabilities such as the Kelvin-Helmholtz instability. \citet{Pittard07} accordingly concluded that the X-ray emission of the wind interaction zone is insensitive to clumping. Still, the clumps penetrating the wind interaction zone trigger turbulent motion that slightly puffs up the wind interaction zone. 

\subsection{Orbital variability \label{theor:variations}}
Using the same code as \citet{SBP} (i.e.\ assuming axisymmetric winds colliding at their terminal velocity), \citet{PittStev} computed synthetic X-ray lightcurves and spectra for three different values of the wind momentum ratio. They considered two mostly adiabatic systems and one radiative case. By construction, these lightcurves were symmetric about phase 0.5. Most of the predicted variations were hence due to the changing optical depth along the line of sight (i.e.\ larger optical depth when the star with the denser wind is in front, decreasing the observed X-ray flux) though occultation by the stellar bodies also played a role.

These results were revised by \citet{PittPark10} who used the 3-D hydrodynamical models of \citet{Pittard09} to predict the variations of the X-ray emission from the wind interaction regions for different cases of O + O binaries. In accordance with \citet{PittStev}, systems with circular orbits were predicted to exhibit phase-dependent X-ray variability only due to changes of the line-of-sight column densities. Since regions emitting X-rays of different energies are predicted to have different spatial extents, the column densities depend on the energy. Overall, in such systems, the maximum X-ray flux is expected at quadrature phases. However, in contrast with the models of \citet{PittStev}, the lightcurves of short-period circular systems as computed by \citet{PittPark10} are not fully symmetric due to orbital aberration. Moreover, in these short-period systems radiative cooling is quite efficient and the models hence predict sharp dips in the soft-band lightcurve when the dense thin layer of cooled post-shock gas crosses the line-of-sight. The latter feature disappears for longer-period systems with adiabatic wind interactions, for which the synthetic lightcurves become more symmetric. 

For eccentric systems, \citet{PittPark10} predict also changes in the intrinsic emission. However, in the systems considered by \citet{PittPark10}, the behaviour is far more complex than the $1/d$ relation expected from Equation\,(\ref{eq:3}). This is because the wind collision region changes from adiabatic at apastron, to highly radiative at periastron. Moreover, the pre-shock velocities also change. As a result, \citet{PittPark10} predict that the soft emission reaches a maximum at periastron ($\phi = 0.0$), whilst the hard emission peaks before, and then drops to almost its minimum value at periastron. Moreover, in the 3-D models of \citet{Pittard09} some dense clumps of cool material formed during periastron passage were found to survive inside the wind interaction region. Because of their high density and inertia, it takes some fraction of the orbital cycle \citep[about half the orbital period in the simulations of][]{Pittard09} to push them out. As a result of the history of the wind interaction zone, the results of \citet{PittPark10} thus reveal strong hysteresis effects with the emission being softer after periastron. These effects are reminiscent of the hysteresis found in the synthetic thermal radio lightcurves computed by \cite{Pittard10}.

Even more complex phenomena have been predicted for eccentric systems as a result of instabilities of highly radiative shocks. In their effort to explain the X-ray lightcurve of WR~22 (WN7 + O, $e \simeq 0.56$, $P_{\rm orb} = 80$\,d), \citet{PG} presented hydrodynamic simulations indicating that the collision region is very unstable if the acceleration of the winds is taken into account. \citet{PG} show that, as the stars approach periastron, the ram pressure of the WR wind overwhelms that of the O-star and following a disruption of the shocks by non-linear thin-shell instabilities, the wind collision region collapses onto the O star.\\

Finally, synthetic X-ray line profiles formed in the hot post-shock gas and their orbital variations were computed by \citet{Henley}. These computations were based on the density and velocity fields from snapshots of 2-D axisymmetric hydrodynamic simulations of adiabatic wind collisions assuming that the winds collide at their terminal velocity. \citet{Henley} focused on profiles of the Ly$\alpha$ transitions of O, Ne, Mg, Si and S which fall into the sensitivity range of the current generation of high-resolution X-ray spectrographs. However these lines are also emitted by the hot plasma intrinsic to the winds of the individual stars that make-up the binary. Therefore, their profiles are likely contaminated by the intrinsic emission of these stars. This is why \citet{RMN} recently computed a series of synthetic line profiles for the Fe {\sc xxv} K line near 6.7\,keV which is much less affected by emission intrinsic to the stars. In their work, \citet{RMN} use the semi-analytical formalism of \citet{Canto} to predict the complicated morphology of the Fe K line complex for different systems and different orbital phases.   

\subsection{Magnetic fields and particle acceleration \label{nonthermal}}
Although magnetic fields have become an important topic in massive stars research, understanding their impact on wind-wind interactions is a subject that is still in its infancy. In the context of colliding winds, magnetic fields are notably responsible for the synchrotron radio emission observed in some of these systems \citep[e.g.][]{PD}. However, the requested field strengths are actually relatively low. For instance, in the case of Cyg\,OB2 \#9 (O5-5.5\,I + O3-4\,III, $e = 0.71$, $P_{\rm orb} = 2.35$\,yr), \citet{Parkin14} estimated that the surface magnetic field required to account for the observed synchrotron radio emission is about 0.3 -- 52\,G. Such values are below the detection capabilities of current spectropolarimetric facilities. The lack of detections of stellar magnetic fields in particle-accelerating colliding wind binaries \citep{CarinaBfield,Neiner} could also indicate that the fields responsible for the synchrotron emission forms in situ, i.e.\ inside the wind interaction zone. 

If we consider that a weak stellar magnetic field exists, what consequences are expected for the wind-wind interaction? \citet{FGA} present the first full magnetohydrodynamic numerical simulation of the wind-wind interaction in $\eta$~Carinae assuming both stars have a weak dipolar magnetic field with a polar field strength of 1\,G. They find that the magnetic field reduces the strength of the shock since part of the upstream kinetic energy is converted into magnetic rather than thermal energy. Behind the shocks, the density of the magnetic field line increases. In an adiabatic case with very large Mach numbers, the magnetic field component parallel to the shock increases by a factor 4. In a radiative shock, the effect is much larger. Yet \citet{FGA} show that the magnetic field evolution at the shock is not linearly correlated to density or thermal pressure. Finally, depending on its orientation, a magnetic field can reduce quite significantly the efficiency of electron conduction \citep{Orlando}.\\ 

Particle acceleration and the subsequent energy losses via synchrotron radiation and inverse Compton scattering are also important ingredients in the physics of colliding winds. Beside the issue of shock modification \citep[][see Sect.\,\ref{shockphysics}]{PD}, the substantial amount of energy required to accelerate particles affects the properties of the shocks and the level of thermal X-ray emission. \citet{PD} present a model of the synchrotron radio emission, the thermal X-ray emission and non-thermal (inverse-Compton) X-ray and $\gamma$-ray emission of WR~140. They argue that the efficiency of diffusive shock acceleration varies with orbital phase and that this could explain why the X-ray emission of this system deviates from the $1/d$ relation (see Sect.\,\ref{excent} below). 

\citet{Reitberger} solved the hydrodynamic equations along with transport equations of energetic particles accounting for diffusive shock acceleration, radiative losses, and adiabatic cooling. In their simulations, none of the electron energy spectra reaches energies above 1\,TeV whilst protons reach energies of several TeV. Orbital motion, although included at a modest level in this simulation, changes the properties of the leading and trailing arms. However, although this is not the subject of the present review, we note that, apart from $\eta$~Car (see Corcoran et al.\ this volume), to date no colliding wind system has been detected at $\gamma$-ray energies. Indeed, \citet{Werner} investigated data from 24 months of {\it Fermi} Large Area Telescope (LAT) observations and failed to detect emission from any out of seven WR + O colliding wind binaries ($\gamma^2$~Vel, WR~70, 125, 137, 140, 146, 147) though they were expected to be good candidates. For two of the stars (WR~140 and 146), the upper limits are well below model predictions of \citet{BR,PD,Reimer06,Reimer07,Reimer09a,Reimer09b}. 

\section{The observational picture \label{observation}}
On the observational side, major progress became possible thanks to the unprecedented sensitivities and spectral resolution of the {\it Chandra} and {\it XMM-Newton} missions. These observatories allowed monitoring the orbital variability of the X-ray emission of interacting wind systems. Thanks to the longevity of these satellites, this became possible also for systems with periods of several years. In addition, both spacecraft have been used to perform large sensitive surveys of clusters and associations hosting massive stars, thereby allowing to refine the global picture of the X-ray emission of massive stars and massive binaries in particular. 

\subsection{The X-ray luminosity}
Using data from the {\it ROSAT} All-Sky Survey, \citet{ber97} showed that the X-ray luminosity $L_{\rm X}$ of O-type stars scales roughly as $10^{-7}\,L_{\rm bol}$, thereby confirming earlier findings with {\it EINSTEIN}. However, there was considerable scatter around this relation. Moreover, the multiplicity of many of the O-stars in the sample was not known, and for many objects only upper limits on the X-ray flux were available.   

Whilst no all-sky survey was performed with neither {\it XMM-Newton} nor {\it Chandra}, both observatories have been used to perform surveys of clusters and/or of large massive stellar populations. These campaigns significantly improved upon the older work. For instance, detections for the full O-star population in a given cluster were available, reducing biases. In parallel, optical spectroscopic monitoring have tremendously improved the binary census, allowing for trends to be better identified.

It must however be noted that details in processing and calibrations as well as assumptions on reddening laws and bolometric corrections render direct comparisons of $L_{\rm X}$\footnote{We note that one must be cautious about the definition of $L_{\rm X}$. Here, the X-ray luminosity will only be corrected for absorption by the neutral interstellar medium. X-ray spectra of massive stars also display absorption due to ionized wind material, but its value critically depends on the signal-to-noise ratio of the data and on the assumptions of the models used by the authors, resulting in several dex differences on the intrinsic luminosity. Therefore, while they could at first be considered as more physical, the meaning of fully dereddened luminosities is ambiguous. The well-known degeneracy between absorption column and temperature in spectral fits (a hot plasma with low absorption providing a fit of equivalent quality to a warm plasma with more absorption) further limits the meaning of ``average'' temperatures \citep[for examples, see][]{naz14magstars}. Unfortunately, both fully dereddened luminosities and average temperatures are still sometimes the only parameters used \citep[e.g.\ ][]{gag12}.} and \lxlb\ values between different papers sometimes difficult \citep[see remarks on a 0.2\,dex change in \loglxlb\ from such simple effects in Sect.\,2.2 of][]{naz11}. 

Nevertheless, all homogeneous studies using new, sensitive data led to the same conclusion \citep{san06,osk05,ant08,naz09,naz11,rau14cyg}: the majority of the known O + OB binaries display \lxlb\ ratios in the same range as supposedly single O stars. This means that not all massive binaries harbour X-ray bright wind-wind collisions; on the contrary, such X-ray bright CWBs appear quite rare, whatever their period \citep{naz09,naz11}. 

Still, on average, binaries have a slightly larger \lxlb\ ratio, by about 0.1--0.2\,dex, especially in harder energy bands \citep{naz09,naz11,rau14cyg}. Although systematic, this difference remains below the observed scatter around the average \lxlb\ relation of supposedly single objects. Hence it is not formally significant and renders the identification of a binary from its sole \lxlb\ ratio rather difficult. 

This lack of strong overluminosities in binaries, apparently independently of the orbital period or the spectral types, remains to be explained. Efficient cooling or radiative braking, as well as reduced mass-loss rates certainly play a role. Most probably, X-ray bright colliding wind binaries populate only a small region of the parameter space. In addition, in some X-ray bright binary systems, the overluminosity is not linked to a wind-wind collision but stems from the presence of magnetically confined winds in one of the stars \citep[e.g. HD~191\,612][]{naz10}. 

To make things more complicated, most of the long-period massive binaries feature highly eccentric orbits and, as a result, the X-ray flux of the colliding winds interaction should vary quite significantly between apastron and periastron. However, as these systems spend most of the time far away from periastron, snapshot X-ray observations are more likely to catch them in a low emission state. It thus appears impossible to draw secure conclusions on binarity from a single X-ray snapshot observation.\\

The situation is more complex for WR stars for which {\it EINSTEIN} and {\it ROSAT} data did not reveal a well defined \lxlb\ relation \citep{Wessolowski}. For WR binaries, recent observations have generally confirmed the earlier results and extended them to the Magellanic Clouds (MCs) \citep{gue08c}: binaries appear brighter than supposedly single stars. This is particularly the case for WC stars, where no single object has been detected in the X-ray domain up to now \citep{osk03,ski06}, with the possible exception of the WC4 star WR~144 \citep{rau14cyg}. For WN binaries, it remains generally true that binaries are brighter than single stars, although there are exceptions where the emission does not seem to come from the wind interaction. For instance, the short-period, eclipsing binary CQ~Cep (WN6 + O9, $P_{\rm orb} = $1.64\,d) displays a constant X-ray emission with typical characteristics of single WN objects hence without obvious traces of colliding wind emission \citep{ski15}. 

A link between X-ray brightness and orbital period was proposed by \citet{Pollock1}, with short-period systems being less bright than those with longer periods, but this has not been checked yet using recent data. \citet{zhe12} further found that even short-period WR + O systems appear to be adiabatic, despite expectations of radiative collisions based on the high density of the WR wind. This problem could be solved if the mass-loss is both clumpy and reduced compared to older evaluations, a scenario which seems very plausible at the present time. Finally, \citet{gue08c} suggest that the detected X-ray emission of WR stars in the MCs is brighter than that of equivalent Galactic objects. If confirmed, this effect could be linked to the lower metallicity in the MCs, which leads to a reduced absorption.\\

Another important point concerns the comparison of the X-ray luminosities predicted by theoretical models with the observed values. In most studies, the comparison between theory and observation does not account for the X-ray emission intrinsic to the individual stars. In some cases, the emission from the wind-wind interaction actually represents only a minor contribution to the full X-ray emission. This is the case of 9~Sgr \citep[O3.5\,V((f$^*$)) + O5-5.5\,V((f)), $e = 0.70$, $P_{\rm orb} = 9.1$\,yr,][]{9Sgr}. Another example is provided by the highly eccentric O9\,III + B1\,III binary $\iota$\,Ori ($e = 0.76$, $P_{\rm orb} = 29.13$\,d). \cite{iotaOri} analysed two {\it ASCA} observations of that system taken at periastron and apastron. The X-ray luminosities and spectral distributions were found to be remarkably similar on both observations, in stark contrast with the expected order of magnitude variation for a colliding winds interaction. This points towards a colliding-wind emission that is unexpectedly weak, and an X-ray emission dominated by intrinsic shocks in the winds. 

Even neglecting the contribution of the intrinsic emission, theoretical models quite frequently overestimate $L_{\rm X}$ by at least an order of magnitude \citep{ZS,ski07,zhe10,zhe12,Montes,zhe15}. As a result of stellar wind clumping, there is nowadays a general concensus that mass-loss rates determined with conventional optical, UV, IR or radio diagnostics might need to be reduced by typically a factor $\sim 3$ \citep[e.g.][]{Puls}. Yet, this reduction is generally not sufficient to bring the theoretical and observed luminosities of colliding wind systems into agreement. In this context, \citet{DeBecker04} analysed an {\it XMM-Newton} observation of HD~159\,176 (O7\,V + O7\,V, $e = 0.0$,  $P_{\rm orb} = 3.367$\,d). Hydrodynamic simulations as well as the semi-analytic approach of \citet{Antokhin04} were used to theoretically predict the X-ray luminosity due to wind-wind interactions. Adopting wind parameters from the literature resulted in a striking discrepancy between the predicted and the observed X-ray luminosities. A significant reduction of the mass-loss rates of the stars was necessary to alleviate part of the discrepancy. The more recent model of \citet{PittPark10} with mass-loss rates of $2 \times 10^{-7}$\,M$_{\odot}$\,yr$^{-1}$ is able to account for the emission level of HD~159\,176, although one should stress that this model again does not account for intrinsic emission. 

In the case of Cyg\,OB2 \#9, \citet{Parkin14} also found that mass-loss rates in their simulations had to be reduced by a factor 7.7 to match the observed level of the X-ray emission.  

In the case of the trapezium-like system QZ~Car which hosts two binary systems with O9.7\,I and O8\,III primaries, \citet{QZCar} noted from simple analytical considerations that the wind interaction in the binary containing the O8\,III star should produce an X-ray emission 10 times stronger than the one observed with {\it Chandra}. \citet{QZCar} suggest that in this system, the wind interaction could be disrupted by mass transfer via Roche lobe overflow from the O8\,III star towards its companion. 

Another striking example is provided by the study of WR~22 by \citet{PG}. These authors used 3-D adaptive-mesh refinement simulations to investigate the dependence of the X-ray emission associated with the wind-wind collision on the acceleration of the stellar winds, radiative cooling, and orbital motion. They accounted for radiative driving, gravity, optically-thin radiative cooling, and orbital motion. The adopted mass-loss rates were $1.6 \times 10^{-5}$ and $2.8 \times 10^{-7}$\,M$_{\odot}$\,yr$^{-1}$ for the WN7 and the O9\,V star, respectively. \citet{PG} showed that when a stable wind collision region exists the models over-predict the X-ray flux by more than two orders of magnitude. They also found that the collapse of the wind interaction zone onto the O star substantially reduces the discrepancy in the 2--10 keV flux to a factor $\sim 6$, though the observed spectra were not well matched by the models. \citet{PG} accordingly suggest that the agreement between the models and observations might be improved by reducing the mass-loss rates of both stars and simultaneously increasing the ratio of the wind momentum ratio to a value close to 500 in favour of the WR star. Under these circumstances, no normal wind ram pressure balance would occur at any orbital phase. In the absence of radiative braking this would lead to a permanent collapse of the wind collision region onto the O star. It remains to be seen whether such drastic changes in the wind parameters are consistent with the observed properties of WR~22 at other wavelength.\\

Beside the overluminosity, another criterion for the presence of X-ray emission due to colliding winds that is frequently used is the spectral hardness. Indeed, fitting the spectra of WR + O binaries with X-ray bright wind-wind interactions requires usually at least two thermal components with $kT$ near 0.6 and 2.--4.\,keV. The harder component is generally subject to higher wind absorption, as expected for colliding winds: X-rays of higher energy being produced closer to the line-of-centers, hence deeper inside the winds. Such a hard component is also sometimes seen in presumably single O-stars, though with a lower strength than in X-ray bright colliding wind systems. More generally, surveys revealed a harder X-ray emission for massive binaries compared to presumably single stars, but the difference is of limited amplitude (see above). Finally, hard X-ray emission is also expected in the case of magnetically confined winds \citep[though it is not always observed,][, ud-Doula \& Naz\'e, this volume]{naz14magstars}, which may blur the picture when searching for colliding wind X-ray emission.

\citet{Raassen} pointed out that prominent colliding wind systems often display a strong Fe K line near 6.7\,keV in their spectra. This line has thus been used as a possible indicator of the presence of wind-wind interactions \citep[e.g.\ in the case of the massive stars in the core of the M\,17 nebula,][]{MR}, although once more it is also seen in some stars with magnetically confined winds. We thus conclude that the presence of a hard X-ray emission is not necessarily or uniquely linked to binarity.

\subsection{Orbital variations}
One of the best ways to confirm the presence of a colliding wind region is through the monitoring of the X-ray emission. Indeed, in most cases, wind-wind interactions should produce phase-locked variations of the observable X-ray flux either because of changing absorption along the line-of-sight as the stars revolve around each other and/or because of the varying separation in eccentric binaries (see Sect.\,\ref{theor:variations}). The only cases were no variations are expected would be for circular orbits seen face-on, which is a statistically rare configuration. 

Variability thus represents a powerful diagnostic of colliding winds, but it requires a set of time-constrained observations, which is demanding in terms of observing time and sometimes challenging in terms of scheduling. Nevertheless, the results are worth the effort and several monitoring campaigns have been performed.
Before, we review the results that have been obtained, let us stress that one can also study variations on other timescales. In particular, \citet{che90} expected ``flares'' occurring when clumps collide but no such short-term events were reported. Stochastic variability on rather short timescales was also expected as a result of dynamical instabilities in radiative wind interaction zones (see Sect.\,\ref{theor:instabilities}), but again, no significant short-term variability of the X-ray emission was found so far \citep[e.g.][]{DeBecker04,san04,Linder}. 

\subsubsection{Orbital variations due to changing absorption and eclipses \label{opacity}}
Photoelectric absorption of the X-rays emitted by the shock-heated plasma occurs primarily in the unshocked stellar winds along the line-of-sight towards the observer. The associated optical depths depend on the wind properties as well as the binary characteristics. When the two stars feature very different wind properties, such as in WR + OB systems, absorption effects may lead to particularly spectacular variations of the X-ray emission.

In this context, the prototypical case is $\gamma^2$~Vel. \citet{Willis} reported a strong increase in the {\it ROSAT} count rate around the time when the O star passes in front: this was readily explained as an effect of the lower absorption of the more tenuous O-star wind filling the shock cone. The width of the peak in the X-ray lightcurve is directly related to the cone opening angle, and in this case indicates a quite narrow shock cone (half opening angle of 20--30$^{\circ}$). Subsequent observations with {\it ASCA} \citep{Stevens96,rau00} and {\it XMM-Newton} \citep{sch04} revealed that the spectrum of $\gamma^2$~Vel is composed of a constant cooler component (interpreted as arising further in the winds), a constant hotter component (which would not be sensitive to absorption changes), and a varying medium-temperature component strongly affected by changes in absorbing column. However, a large reduction in mass-loss rate was required to fit the observed absorption value \citep{sch04}.

The lightcurve of V\,444~Cyg also displays a spectacular modulation in the soft band (0.4--2.\,keV), with a peak occurring after the O-star conjonction and a decline steeper than the rise \citep{lom15}. This peak can be explained by three factors: the smaller absorption by the WR wind as the line-of-sight avoids the densest parts of the WR wind as the shock cone wrapped around the O-star turns towards the observer, the lower absorption by the more tenuous O-star wind (filling the shock cone) entering into the line-of-sight, and the addition of the O-star's intrinsic X-ray emission. Note that a simple model considering only absorption by the WR wind is already able to explain the presence of a peak in the soft band. The shift of the soft emission peak with respect to conjunction time is linked to the emitting regions not being exactly centered on the binary axis, suggesting aberration of the shock cone by Coriolis deflection. The asymmetry of the peak can then be reproduced by considering a brighter emission on the leading side of the collision zone. Additional evidence for the role of Coriolis deflection in this system comes from  the secondary eclipse of the hard X-ray emission that is not centered on the time of optical eclipse\citep{lom15}. Finally, the wide shock opening angle inferred from the {\it XMM-Newton} data indicates that radiative braking must be at work in this system \citep{lom15}, as proposed by \citet{Gayley}.

In WR~25 (WN6ha + O4f, $e \sim 0.5$, $P_{\rm orb} = 208$\,d), the flux at energies below 2\,keV rises as the stars move towards periastron (separation effect, see below) but this rise is abruptly stopped by a sharp decline occurring as the WR star passes in front, nearly at the same time as periastron \citep{pan14}. The lightcurve then slowly recovers, as the WR wind slowly quits the line-of-sight. 

In WR~140, the situation is similar: a gradual increase of the flux as the stars move towards periastron is interrupted by a sharp decrease due to absorption when the O star passes behind the WR as seen from Earth \citep{cor03,cor11}. The lightcurves appear compatible with that expected from a point source at the stagnation point rotating with the binary period \citep{cor11}, but the observed value of absorption remains lower than predicted \citep{Pollock}. Other examples of sharp drops in the X-ray flux when the WR wind comes into view include WR~21a \citep[WN + O, $e = 0.7$, $P_{\rm orb} = 32$\,d,][]{Gosset15} and WR~22 \citep[WN7ha + O, $e \sim 0.6$, $P_{\rm orb} = 80$\,d,][]{gos09}. In the latter case, the observed value of the absorption remains much below the predicted value \citep{gos09}.

Two examples of absorption effects have also been found in the Magellanic Clouds. HD~5980 (LBV + WR, $e = 0.3$, $P_{\rm orb} = 19.3$\,d) was the first extragalactic massive binary where X-ray emission from colliding winds was securely identified \citep{naz07}. The system displays a clear increase in observed flux when the star with the less dense wind passes in front \citep{naz07}. The triple system HD~36\,402 (WC4 + O + O with orbital periods 3\,d and 4.7\,yr, though the orbits are not well known) displays a hard, bright, and variable X-ray flux compatible with the 4.7\,yr period \citep{wil13}. The system appears X-ray faint when it is IR bright. This may be linked to the periastron passage as it is the time when episodic dust formation occurs. The wind interaction region would then be deep inside the WR wind and the resulting strong absorption would extinguish the X-ray emission.

Beside WR systems, another example of the effect of changing wind attenuation is provided by the non-eclipsing binary HDE~228\,766 ($P_{\rm orb} = 10.7$\,d, $e = 0$). This system hosts an O8\,II primary and a secondary star which is probably in an intermediate evolutionary stage between an Of supergiant and an WN star. \citet{Rauw14} analysed {\it XMM-Newton} spectra at both conjunctions and at one quadrature phase to measure the variations of the line-of-sight absorption toward the X-ray emission from the wind-wind interaction zone and thus infer information on the wind densities of both stars. A toy model based on the formalism of \citet{Canto} was used to set constraints on the stellar wind parameters by attempting to reproduce the observed variations of the relative fluxes and wind optical depths at 1 keV. The variations of the X-ray spectra suggest an inclination in the range 54 -- 61$^{\circ}$ and a value of $\eta < 0.2$.

Examples of the effect of changing absorption along the line-of-sight can also be found among O + OB binaries. For instance, \citet{Arias} showed that the {\it ROSAT}-PSPC lightcurve of HD~165\,052 (O6.5\,V + O7.5\,V, $e = 0.09$, $P_{\rm orb} = 2.95$\,d) displays a double-peaked structure with maxima at quadrature phases in agreement with expectations from theoretical models for binaries with roughly equal wind strengths \citep{PS}. Another case is probably the triple system HD~150\,136 consisting of an O3\,V((f$^*$))-3.5\,V((f$^+$)) primary and an O5.5-6\,V((f)) secondary on a circular orbit (P$_{\rm orb} = 2.67$\,d) and an O6.5-7\,V((f)) tertiary on a much wider orbit \citep{Mahy}. \citet{ski05} presented a 90\,ks {\it Chandra} observation of this system which displays variability on a timescale of less than one day. When folded with the ephemerides of \citet{Mahy}, the lightcurve yields a minimum at phase 0.0, i.e.\ at the conjunction phase with the primary component in front. 

However, the detection of such a modulation is not always a proof for the presence of additional X-ray emission coming from the interacting wind region. Indeed, in the case of the contact binary 29~CMa (O8\,Iaf + O9.5\,I, $e = 0$, $P_{\rm orb} = 4.39$\,d) the {\it ROSAT}-PSPC lightcurve shows a roughly sinusoidal modulation with a maximum when the secondary star is in front and a minimum when the primary is in front. \citet{BS} suggested to explain this result through the sole effect of the wind opacities on the intrinsic emission of each compoent without the need to invoke additional X-ray emission from a wind-wind interaction.\\ 

There is also at least one counterexample where a modulation of the X-ray emission due to a changing line-of-sight optical depth is expected, but is not observed: In WR~79 (WC6-7 + O5-8, $e = 0$, $P_{\rm orb} = 8.9$\,d), the X-ray emission in an {\it XMM-Newton} monitoring was found to be constant despite strong expected changes in this asymmetric system \citep{gos11bsrsl}. High spatial resolution {\it Chandra} data revealed the presence of two nearby sources, within the {\it XMM-Newton} PSF hence affecting the {\it XMM-Newton} results \citep{gos11bsrsl,zhe12}. However, while any variability analysis should be considered as preliminary because of that contamination, it would be an extraordinary coincidence that a change in WR~79 would exactly be compensated by a change in those sources in the {\it XMM-Newton} monitoring: the observed constancy is thus more secure, but is difficult to interpret in the context of colliding winds.\\ 

Provided the orbital inclination is right, one could expect eclipses of the wind interaction zone. Indeed, the strongest, head-on wind collision occurs at the so-called stagnation point, located on or near the binary axis. This central region is regularly occulted by the stellar bodies in high-inclination systems. This is the case of V\,444~Cyg where monitoring with {\it XMM-Newton} revealed the presence of eclipses in the hard ($> 2$\,keV) band \citep[][see above]{lom15}. 

In WR~20a (WN6ha + WN6ha, $e = 0$, $P_{\rm orb} = 3.7$\,d), the two stars are identical hence no absorption change is expected but the system undergoes photometric eclipses because of its high inclination \citep[about 75$^{\circ}$,][and references therein]{rau05} hence one could also expect X-ray eclipses. However, a set of {\it Chandra} observations showed no eclipses in the X-ray domain, on the contrary, the system appeared brighter at conjunction phases than at phases just before \citep{naz08}. The absence of eclipses can only be understood if the X-ray emission zone is large while the fainter emission near quadrature could be explained by (self-)absorption in the shocked winds \citep{naz08}. Hydrodynamic models of the system by \citet{Montes} confirmed these conclusions. Because of the small separation between the stars, the winds collide while they are still accelerating. Along the binary axis, the winds thus collide at low speed and the innermost part of the wind interaction region is most probably highly radiative \citep{Montes}. The plasma there is thus likely too cold to contribute significantly to the X-ray emission. The main contribution to the X-ray emission thus arises from the outer parts of the wind interaction zone and this could explain the absence of eclipses in X-rays as found by \citet{naz08}. The models of \citet{Montes} also predict sharp decreases in the X-ray fluxes at quadrature phases because of the large column density along the contact discontinuity. However, these simulations neglect the effect of Coriolis forces which bend the interaction region hence smearing out the events predicted by \citet{Montes} and leading to a rather broad maximum of the column density \citep{Lemaster}. 

In the eclipsing binary CPD$-41^{\circ}$\,7742 (O9\,V + B1-1.5\,V, $e=0.02$, $P_{\rm orb} = 2.44$\,d) {\it XMM-Newton} EPIC data reveal a peculiar lightcurve. The X-ray flux is constant at a low level over half of the orbit when the rear side of the B-star is seen. The X-ray flux then increases except for a clear drop that matches the optical secondary eclipse \citep{CPD7742}. This lightcurve is interpreted as the signature of the primary wind crashing onto the photosphere of the secondary or, in case radiative inhibition and radiative braking are efficient, a wind interaction zone located very close to the surface of the secondary. This situation generates a hot plasma close to the photosphere of the B-star and an effect similar to lunar phases explains the observed gradual rise and decline in flux \citep[see][]{CPD7742}. 

\subsubsection{Orbital variations due to changing separation \label{excent}}
In addition to the changing line-of-sight column densities, eccentric binaries can also display changes of the intrinsic colliding wind X-ray emission because of the changing separation $d$ between the stars. For fully adiabatic wind interactions, the separation between the stars is usually high and the winds should thus collide with velocities close to $v_{\infty}$ at all orbital phases. Therefore, Equation\,(\ref{eq:3}) predicts variations of the X-ray flux as $1/d$. This results in a brightening as the stars move towards periastron, and a decrease of the X-ray flux as the stars separate again. This effect should be most prominent at higher energies, as these photons are less affected by absorption. In systems with purely radiative wind interaction zones, the interaction occurs at lower wind speeds around periastron, therefore leading to a softer and also fainter emission. 

\begin{figure}
\begin{center}
\includegraphics*[width=0.48\textwidth,angle=0]{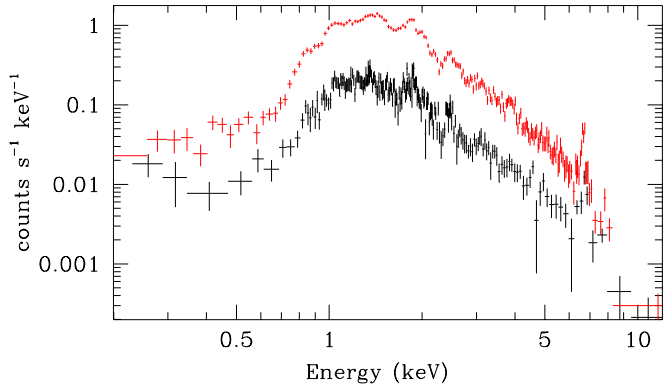}
\end{center}
\caption{Variation of the {\it XMM-Newton} EPIC-pn spectrum of Cyg\,OB2 \#9 between $\phi = 0.190$ (black data points) and $\phi = 0.997$ (red data points). Note the strong variation in flux level between the two observations, the one near periastron being about a factor 4 higher in good agreement with the $1/d$ variation between both observations \citep {YN12}.\label{figure3}}
\end{figure}

Probably one of the best examples of an adiabatic wind-wind interaction displaying a $1/d$ variation is provided by the non-thermal radio emitter Cyg\,OB2 \#9 (see Fig.\,\ref{figure3}). \citet{YN12} present {\it XMM-Newton} and {\it Swift} observations collected around the 2011 periastron passage. The only slight departure from the $1/d$ relation concerns the hard X-ray flux which falls slightly short of the expected value at periastron. \citet{YN12} suggest that this may be due to the collision becoming slightly radiative as a result of the increased wind density at periastron. \citet{YN12} also found that the emission was somewhat softer near periastron and suggested that this might hint at the effects of radiative inhibition or braking \citep{SP,Gayley} becoming more efficient as the separation between the stars decreases. This assertion was confirmed by detailed 3-D adaptive mesh refinement hydrodynamic simulations of Cyg\,OB2 \#9 accounting for wind acceleration, radiative cooling, and the orbital motion of the stars presented by \citet{Parkin14}. In these simulations, wind acceleration was found to be inhibited at all phases by the radiation field of the opposing star. \citet{YN12} showed that the X-ray emission of this system is intrinsically quite hard ($kT > 1.5$\,keV) as expected for an adiabatic wind interaction zone. This hot shocked plasma of the wind interaction zone was also found to contribute significantly to the free-free radio emission of the system \citep{Blomme13}.

The non-thermal radio emitter 9~Sgr is another example of a highly adiabatic wind interaction zone. However, \citet{9Sgr} found that the hard (2.0 -- 10.0\,keV) X-ray flux of 9~Sgr at periastron is lower than expected from a simple $1/d$ scaling relation. Since the stars have a weaker radiation field and are much farther apart from each other than in Cyg\,OB2 \#9, \citet{9Sgr} argue that radiative inhibition or braking are unlikely to play a role. They suggest instead that shock modification by relativistic ions \citep{PD} might be at work in this system.     

Other systems that show hints for a $1/d$ variation of their X-ray flux, at least at high energies, are HD~93\,403 \citep[O5.5\,I + O7\,V, $e = 0.23$, $P_{\rm orb} = 15.1$\,d,][]{HD93403}, HD~93\,205 \citep[03\,V + O8\,V, $e = 0.37$, $P_{\rm orb} = 6.1$\,d,][]{Igor} and WR~25 \citep{pan14}. In HD~93\,403 and WR~25, the variations in the softer energy bands are contaminated by absorption effects. Moreover, at least in the case of HD~93\,403, the offset due to the intrinsic emission of the stars yields changes of the X-ray flux that are lower than expected from a simple $1/d$ scaling.

However, the predicted $1/d$ variations are not always observed. In WR~22 and $\gamma^2$~Vel, the hard part of the spectrum remains constant despite the large variation in separation due to a high eccentricity \citep{gos09,rau00,sch04}. \citet{PG} presented tailored hydrodynamic models of WR~22. Both instantaneously accelerated and radiatively driven stellar winds were considered and compared against the observations of \citet{gos09}. Neither of the models actually reproduced the observed lightcurve well.

In the same context, repeated monitoring campaigns of WR~140 prooved the existence of variations of its intrinsic flux. However, these variations by about a factor three remain well below those expected (a factor 19) from the $1/d$ relation for such a large eccentricity \citep{pol95,ZS,cor11,sug11}. As pointed out above, \citet{PD} suggested that, in the case of WR~140, this situation results from the impact of particle acceleration on the properties of the shocks.\\

For shorter period eccentric systems, there are strong deviations from the $1/d$ relation and the wind interaction zone can switch from mostly adiabatic near apastron to highly radiative near periastron. As a result, theoretical models of \citet{PittPark10} predict that the X-ray flux at a given orbital phase depends on the history of the plasma heated at earlier orbital phases. There are now a number of systems where these predictions have been verified by observations. To identify such a behaviour requires a rather good phase coverage of the orbital cycle. For instance, \citet{Cazorla} analysed a sample of {\it XMM-Newton} and {\it Swift} observations of the core of the Cygnus\,OB association, significantly improving the phase coverage of the 21.9\,d eccentric O6If + O5.5III(f) binary Cyg\,OB2 \#8a ($e = 0.21$). The X-ray emission clearly displays phase-locked variations that are anticorrelated with the changes of the non-thermal radio emission of this system. In a system such as Cyg\,OB2 \#8a, where the wind interaction zone is at least partially radiative, the collision occurs at higher speeds at apastron, when the winds have enough room to accelerate. This leads to a stronger shock at apastron, and thus a harder X-ray emission. However, the X-ray flux does not peak at apastron, but at an intermediate phase of $\phi \sim 0.8$. A plot of the X-ray flux versus the orbital separation \citep[Fig.\,3 of][]{Cazorla} revealed striking similarities with Fig.\,19 of \citet{PittPark10}. The predicted hysteresis behavior is clearly seen with the emission being harder at phases when the stars get closer than when they separate again. 

\begin{figure*}
\begin{center}
\includegraphics*[width=0.45\textwidth,angle=0]{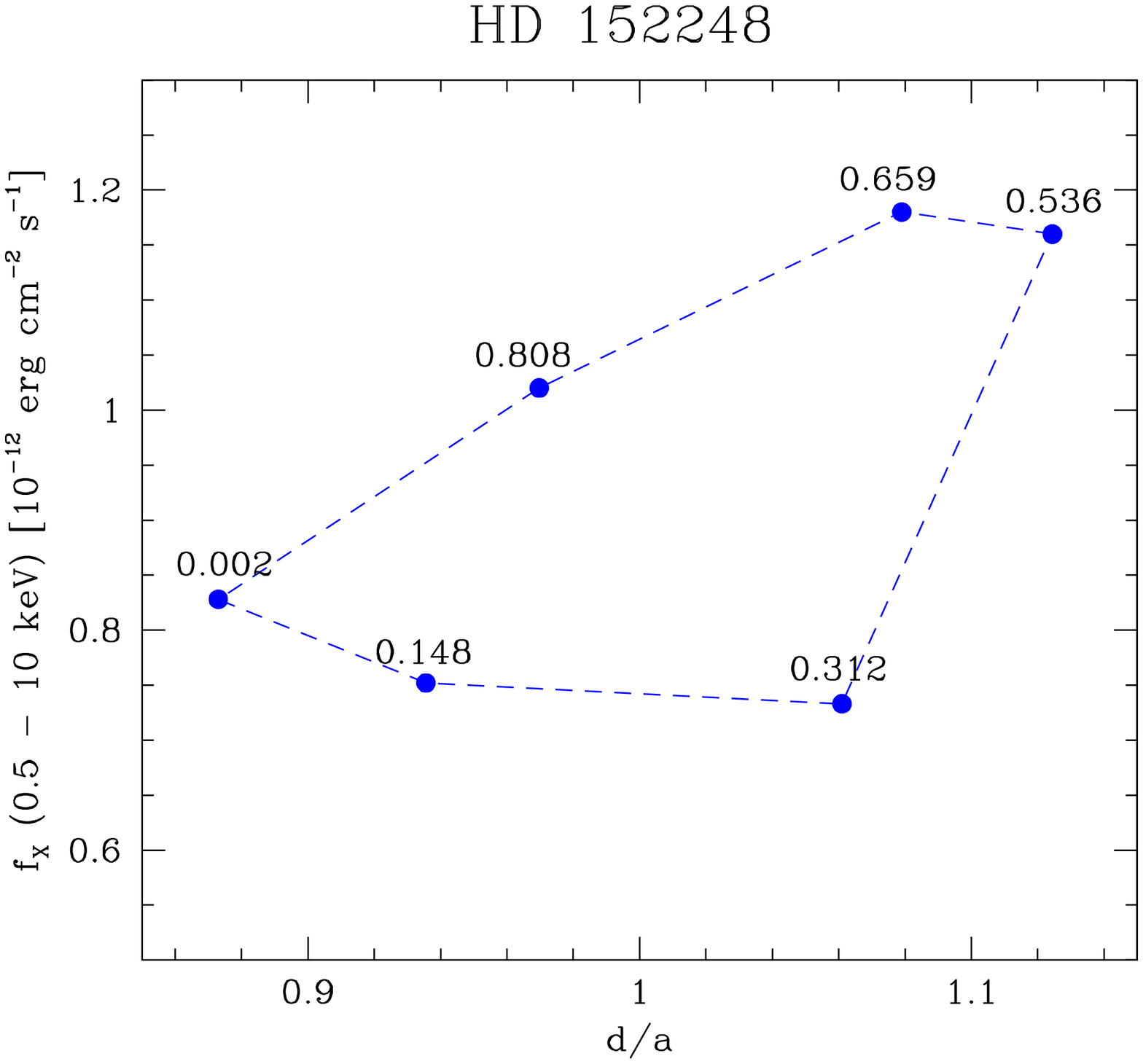}
\includegraphics*[width=0.45\textwidth,angle=0]{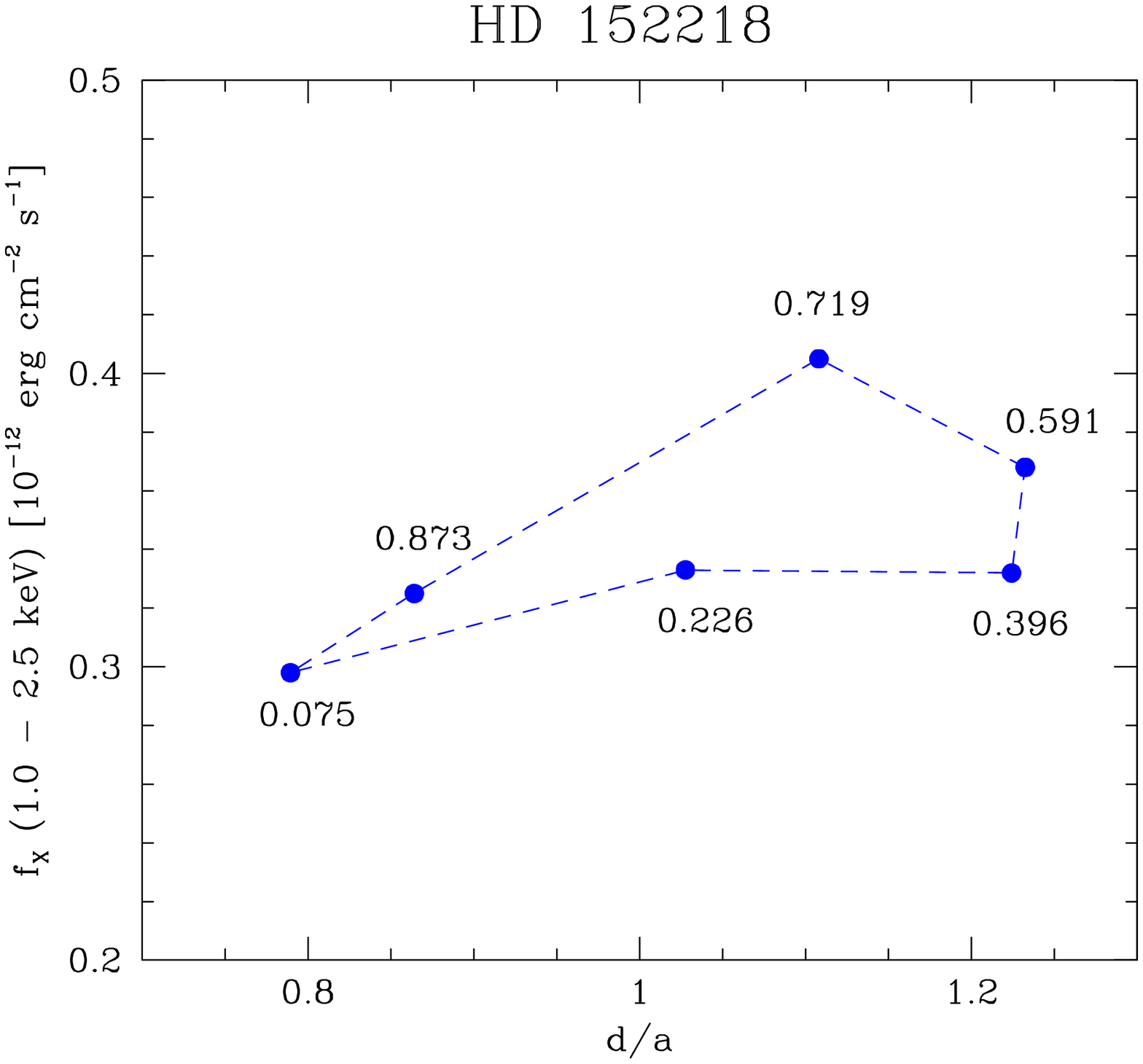}
\end{center}
\caption{Left: total X-ray flux of HD~152\,248 observed with {\it XMM-Newton} as a function of orbital separation \citep{san04}. Right: same for the medium-band flux of HD~152\,218 \citep{HD152218}. The hysteresis-like behaviour is clearly seen in both cases. The orbital phases of the observations are indicated by the labels.\label{figure2}}
\end{figure*}

Similar effects are present in HD~152\,248 (O7.5(f)\,III + O7(f)\,III, $e = 0.13$, $P_{\rm orb} = 5.816$\,d) and HD~152\,218 (O9\,IV + O9.7\,V, $e = 0.26$, $P_{\rm orb} = 5.60$\,d) as illustrated in Fig.\,\ref{figure2}. \citet{san04} found the X-ray flux of the highly radiative system HD~152\,248 to display a clear, asymmetric modulation with orbital phases, reaching a maximum at $\phi = 0.66$. In the case of HD~152\,218, \cite{HD152218} also found that the X-ray flux and its hardness increase near apastron.

Yet another example of an hysteresis is found in the X-ray lightcurve of WR~21a (WN + O, $e = 0.7$, $P_{\rm orb}  = 32$\,d). In this system an increase of the flux is detected towards periastron, but the flux variations are clearly dependent on both separation and phase \citep{Gosset15}. Furthermore, just before periastron, the X-ray flux suddenly decreases. This change in behaviour cannot be due to the sole effect of an increasing absorption as it is also seen at hard energies which are far less affected by photoelectric absorption. This situation instead suggests a crash of the wind interaction zone onto the surface of the O-star \citep{Gosset15}.\\ 

There are a number of systems where X-ray observations suggested the presence of wind interactions, but accurate orbital elements are so far lacking. This is the case of the triple or quadruple system Cyg\,OB2 \#5 \citep{Cazorla}. The variations of the X-ray flux indicate that they do not arise from the short-period (6.6\,days) O6.5-7\,I + OB-Ofpe/WN9 inner binary, but rather from the collision between the winds of the inner binary and that of the tertiary component orbiting it with a 6.7\,yr period as derived from the modulation of the non-thermal radio emission \citep{Kennedy}. Yet, there are currently huge uncertainties on the orbital elements of the tertiary which prevent a more thorough interpretation \citep{Cazorla}. The same is true for WR~48a \citep{zhe14,zhe14b} and WR~65 \citep{osk08}: variations of the X-ray emission have been observed, but are difficult to interpret without a detailed knowledge of the orbit. 

In other systems, the scarce number of data prevents a detailed investigation. For instance, {\it XMM-Newton} data of WR~137 suggest some variations compared to the {\it Einstein} or {\it ROSAT} observations though all observations were taken at similar separations \citep{zhe15}. 

\subsection{Observational insight into shock physics}
As pointed out in Sect.\,\ref{shockphysics}, in wide systems, the shock-heated plasma could be out of ionization equilibrium and the electron and ion temperatures are likely non-equal. For WR~140, \citet{ZS} found that models with non-equal electron and ion temperatures were favored. This finding was confirmed by \citet{Pollock} on the basis of theoretical considerations (at apastron, the Coulomb interaction length is similar to the orbital separation), but also from the fact that line width seems to increase with ionization potential. This latter result suggests that Ne and Mg lines are produced more centrally (closer to the binary axis where the spread in velocities is expected to be smaller) than the Si or S lines. In ionization equilibrium, one would rather expect the hottest plasma (hence emission lines associated with ions of higher ionization potential) to arise closer to the binary axis. A similar problem is present when comparing the width of He-like triplets to lines of H-like ions, again suggesting that they do not arise from the same region. \citet{Pollock} thus concluded that ionization balance is not reached. These authors also reported the simultaneous presence of lines from many different ionization stages of iron (Fe\,{\sc xvii} -- Fe\,{\sc xxv}) in the {\it Chandra}-HETG spectra between 10.6 and 15.0\,\AA. \citet{Pollock} suggested that there is no efficient electron heating of the pre-shock gas because of the magnetic field at the shock (WR~140 is a non-thermal radio emitter) but no detailed calculations were presented. 

In a similar manner, \citet{Zhekov} suggested that, in view of timescales, ionization equillibrium should not be reached in the very wide system WR~147 (WN8 + OB). He noted that non-equilibrium ionization, collisionless shock models give a better fit to the data. 

For $\gamma^2$~Vel, the scaling relations indicate that the shocked plasma should be in ionization equilibrium. However, the analysis of {\it Chandra} data by \citet{hen05} indicated that Mg lines are formed in hotter plasma located closer to the O-star than Si lines. This is a typical signature of non-equilibrium ionization. This result was however not recovered in the analysis of \citet{ski01} of the same data, as the authors used different calibration files and background subtraction: the presence of non-equilibrium ionization in $\gamma^2$~Vel remains thus currently an open issue.\\

A very peculiar situation concerns the wind interaction in Plaskett's Star (HD~47\,129, O6\,I + O7.5\,I, $e = 0$, $P_{\rm orb} = 14.4$\,d). Combining {\it XMM-Newton} observations with {\it ROSAT} data, \citet{Linder} found a modulation of the X-ray emission with orbital phase with a minimum when the primary star is in front, although there was quite substantial scatter in the lightcurve. This suggests quite a strong occultation effect by the body of the primary star in this non-eclipsing binary and thus implies that the wind interaction zone is located closer to the surface of the primary than one would expect from the wind momentum balance. Optical spectra of HD~47\,129 presented by \citet{Linder2} revealed an annulus of circumstellar material, likely due to a confinement of the wind of the secondary. The primary wind would then collide with this confined wind, leading to strong X-ray emission at the contact between the confined secondary wind and the incoming primary wind and shifting the interaction zone closer towards the primary than one would expect without confinement of the secondary wind. Whilst \citet{Linder2} attributed this confinement to the fast rotation of the secondary, subsequent spectropolarimetric observations by \citet{Grunhut} led to the discovery of a magnetic field of dipole polar strength $2.85 \pm 0.5$\,kG, suggesting that this magnetic field is responsible for the confinement of the secondary's wind. Therefore, Plaskett's Star likely features a complex combination of X-ray emission from the shocks within the secondary's wind as it is confined by the magnetic field as well as from the collision between the primary wind and the magnetosphere of the secondary. To date, this is the only system known to exhibit such a configuration.

\subsection{High resolution X-ray spectroscopy}
{\it Chandra} and {\it XMM-Newton} have initiated the era of high-resolution X-ray spectroscopy for moderately bright X-ray sources. For massive stars, HETG and RGS spectra unveiled for the first time some details of the morphology of the spectral lines \citep[see e.g.][for a review]{GN}. 

Only few interacting wind systems have been observed at high spectral resolution in X-rays. This is mostly due to the limited sensitivity of current facilities, but clearly this is a promising route for future observations. Indeed, important information can be derived from high-resolution spectra, including the chemical composition of the emitting plasma, the position of the emission region (through the analysis of He-like triplets), and the velocity of the emitting plasma (through line width and shift). For example, in single massive stars, the forbidden lines of He-like triplets are suppressed through UV pumping of the upper level of the transition as the X-ray emission occurs in a region close to the star, hence flooded by intense UV radiation. Wind-wind collisions occur farther away from the photospheres and the forbidden lines therefore remain strong (e.g. WR~147, \citealt{zhe10b}; WR~48a, \citealt{zhe14}; WR140, \citealt{Pollock}; $\theta$\,Muscae, \citealt{sug08}; $\gamma^2$\,Vel, \citealt{ski01,sch04}), which is a strong argument in favour of X-ray emission arising inside the colliding wind region.\\ 

Furthermore, while detailed line profile analysis is not yet possible, the line shift and width have been recorded in a few cases. Widths are generally in the 1000 -- 2000\,km\,s$^{-1}$ range \citep{Pollock,sug08,zhe10b,zhe14}, i.e.\ of the same order of magnitude as wind speeds, as expected. Line shifts are directly related to the orientation of the colliding wind zone. For example, in the triple system $\theta$~Mus (WC5-6 + O6-7\,V + OB\,I, $P_{\rm orb} = 19$\,d), lines appear redshifted \citep{sug08}: if the orbital solution of the close binary is correct, then these redshifted lines cannot be associated with a shock cone within the binary, folded around its O component because its orientation at that time would rather yield blueshifted lines. Instead, these redshifts would arise in a collision between the combined wind of the binary and the wind of its optical companion, an O-type supergiant \citep{sug08}. Still, the ephemeris must be checked and the line shifts must be monitored closely before definitive conclusions can be drawn in this case. 

A very special situation is provided by the very wide system WR~147. To date, this is the only example where the spatial extension and location of the X-ray source were measured using high-spatial resolution {\it Chandra} data. First evidence for spatially extended emission associated to a wind interaction zone was reported by \citet{pit02}. Subsequent deeper observations by \citet{zhe10} revealed further details: a northern source which is extended and coincident with the bowshock observed at radio wavelengths, and a southern one that coincides with the position of the WN8 star. In view of its location and spectrum, the former component was therefore considered as the signature of the wind-wind collision between the WR and the OB companion. However, the latter component displays a bright, hard, and variable emission untypical of late WN stars: \citet{zhe10b} therefore proposed that the WR actually is a close binary where a second wind-wind collision takes place. In WR~147, the lines from the northern component appear blueshifted, indicating a colliding wind zone located between the observer and the WR \citep{zhe10b}. However, owing to its very long period, the orbit of WR~147 is essentially unknown. 

The orbit of WR~140 is far better determined and the two high-resolution spectra presented by \citet{Pollock} reveal line variations clearly associated to the geometry of the shock zone. Before periastron, the shock cone is opened towards us and the lines naturally appear blueshifted; after periastron, the cone has turned away and the lines are thus slightly redshifted (and somewhat broader). 

In the poorly-known episodic dust maker WR~48a, spectral lines seen in a high-resolution {\it Chandra} spectrum are also blueshifted but this result is more difficult to interpret because of the absence of accurate orbital elements \citep{zhe14}. In fact, the {\it Chandra} observation indicates a larger absorption than what is measured on an {\it XMM-Newton} observation of WR~48a \citep{zhe11,zhe14}, suggesting the WR-star was in front of its companion at the time of the {\it Chandra} observation. The observed blueshift suggests that the cone is wrapped around the WR star (or the star in the system with the wind producing the largest absorption). This implies that the companion should have a more energetic wind, therefore suggesting that WR~48a is a WR + WR system, instead of a classical WR + O binary \citep{zhe14}. This assumption was confirmed by subsequent optical and near-IR observations that revealed a WC8 + WN8h system \citep{zhe14b}. 

In $\gamma^2$~Vel, no line shift was detected on a {\it Chandra} grating observation taken near X-ray maximum \citep{ski01,hen05}, which is at odds with expectations. At that phase, one expects the shock cone to be opened towards Earth, which is demonstrated by the higher apparent flux linked to the lower absorption of the O-star wind (see above). To get a negligible line shift under these conditions, the only possibility is to have a wide opening of the shock cone \citep[half opening angle $>80^{\circ}$][]{hen05}. This could be due to radiative braking by the O-star's radiation, but a wide cone is incompatible with the X-ray lightcurve: the short duration of the low absorption episode, occurring when the line-of-sight crosses the wind of the O-star clearly requires a narrow opening \citep[half opening angle $\sim25^{\circ}$]{Willis,hen05}. A possible solution to the problem could be contamination of the spectral lines by the intrinsic X-ray emission of the O-star which should be essentially unshifted.\\ 

The X-ray spectra of several carbon-rich Wolf-Rayet binaries exhibit radiative recombination continua (RRCs) mainly associated with carbon. This is the case for $\gamma^2$~Vel \citep{sch04} and $\theta$~Mus \citep{sug08}\footnote{In addition, hints of an RRC in a lower-resolution {\it Suzaku} CCD spectrum of WR~140 were reported by \citet{sug11}, although such detections are less secure than those with high-resolution spectrographs.}. For the latter system, this feature appears redshifted. The presence of RRCs requires the presence of low-temperature gas that is ionized by the X-ray emission (probably from the colliding wind interaction). This cool gas could be associated with material downstream of the shock cone. 

\section{Conclusions and future perspectives \label{conclusion}}
Over the past fifteen years, there has been tremendous progress in our understanding of wind-wind interactions in early-type binaries. Theoretical models with an increasing degree of sophistication have been developed and observations with the current generation of X-ray observatories allowed to study the phenomenon with unprecedented details over a wide range of the parameter space. In particular monitoring of interacting wind systems with low-resolution CCD spectroscopy has revealed the complexity of the X-ray lightcurves as well as their energy dependence. Some of the observational results, such as the $1/d$ flux modulation in some eccentric adiabatic systems and the hysteresis in shorter-period eccentric systems, provide nice confirmation of the theoretical predictions. Other findings, such as the lack of $1/d$ variations in some wide eccentric systems or the lower than expected level of the colliding wind X-ray emission, still challenge the theory. An important question to be addressed in future studies concerns a better understanding of the ionization conditions (equilibrium versus non-equilibrium) of the post-shock plasma. Another aspect to be investigated is why some massive binary systems are X-ray dim whereas others are prominent X-ray emitters. To address the latter question one needs to consider binary systems over as wide a range as possible of the relevant parameters (orbital separations, eccentricities, mass-loss rates, wind velocities,...). In addition to X-ray monitoring campaigns, such studies also require a major observational effort in the optical domain to establish accurate orbital parameters.   

The coming generations of X-ray observatories, especially the JAXA-led mission {\it Astro-H} \citep{AstroH} and the ESA-led observatory {\it Athena} \citep{Nandra}, will enable even finer observational studies. In fact, these observatories will carry microcalorimeter spectrographs, called the Soft X-ray Spectrometer \citep[SXS,][]{SXS} for {\it Astro-H} and the X-ray Integral Field Unit \citep[X-IFU,][]{XIFU,Ravera} for {\it Athena}. These instruments will have unprecedented spectral resolution around the 6.7\,keV Fe {\sc xxv} K complex, allowing for the first time to resolve the line profiles and follow their changes. For early-type stars, the Fe K line is only detected in the spectra of colliding wind binaries (e.g.\ Fig.\,\ref{figure3}) or of stars with magnetically confined stellar winds. It is thus an efficient tracer of the presence of very hot plasma, which - in the case of wind-wind collisions - arises in the innermost part of the interaction zone. Phase-resolved observations of the Fe K lines for a sample of massive binaries (with circular or eccentric orbits) will hence provide new insight into the physics of the phenomenon \citep{SR,RMN}. {\it Athena} will also allow us to enlarge the sample of extragalactic colliding wind binaries that can be studied in X-rays. This is especially interesting in terms of the theoretically predicted dependence of the mass-loss rates on metallicity. Such a dependence could affect both the intrinsic level of the X-ray emission of the wind-wind interaction zone and the level of photoelectric absorption by the unshocked winds. 

\section*{Acknowledgements}
The authors acknowledge support through an ARC grant for Concerted Research Actions, financed by the Federation Wallonia-Brussels, from the Fonds de la Recherche Scientifique (FRS/FNRS), as well as through an XMM PRODEX contract (Belspo). We thank the referees for constructive reports that helped us improve our manuscript, as well as Drs.\ E.R.\ Parkin and E.\ Gosset and the editors of A\&A for granting us the permission to reproduce a figure from their work.

%\bibliographystyle{elsarticle-num-names}
%\bibliography{cwb}

\end{document}